\documentclass[11pt]{article}
\usepackage{moriond,epsfig}
\usepackage{amssymb}
\usepackage{marvosym}
\def\be{\begin{equation}}
\def\ee{\end{equation}}
\def\bea{\begin{eqnarray}}
\def\eea{\end{eqnarray}}

\begin{document}
\vspace*{4cm}
\title{STUDIES OF BEAUTY AT H1 AND ZEUS}

\author{ I. BLOCH }

\address{DESY, ZEUS Group, Notkestr. 85, \\
D-22607 Hamburg, Germany}

\maketitle\abstracts{
    Recent measurements of beauty production in ep collisions at HERA are 
    presented. The results were obtained using data from the experiments H1 
    and ZEUS from the years 1996 to 2000 (HERA I). An outlook is given for 
    analysis on data from HERA II. Differential measurements using muons and 
    jets in deep inelastic scattering and photoproduction are compared to NLO
    QCD predictions. The beauty contribution to the proton structure function
    F$_2$ was determined at large $Q^2$ ($Q^2 > 150~\textrm{GeV}^2$) and is 
    compared with predictions.
}

\section{Introduction}
At HERA (DESY, Hamburg) electrons (positrons) are brought into collision with
protons at a centre-of-mass energy of 318~GeV (300~GeV up to 1997). From 1996
to 2000 the two experiments H1 and ZEUS collected about 
$100~\mathrm{pb^{-1}}$, on which the results presented here are based.
The measurement of beauty production still is a rich testing ground for QCD,
as several hard scales can be chosen, depending on the process to be 
described. The beauty mass always provides one hard scale, but due to 
the presence of additional scales, calculations are not straight forward.
In HERA ep collisions, beauty quarks are predominantly produced in boson gluon 
fusion; a photon from the lepton and a gluon from the proton 
collide to produce a $\mathrm{b\bar{b}}$ pair. This process introduces 
several scales relevant for the calculation of the cross sections:
  \begin{center}
    \begin{tabular}[hb]{llll}
      mass of the b quark                & $m_b$ & $\sim 5~\mathrm{GeV}$  &\\ 
      transverse momentum of the b quark & $p_T^b$&$\sim$ typically&a few GeV \\
      virtuality of the exchanged photon & $Q^2$&$ \lesssim 1~\mathrm{GeV}^2$ &$\equiv\mbox{Photoproduction ($\gamma$p})$\\
                                         &     &$\gtrsim 1~\mathrm{GeV}^2$ &$\equiv \mbox{Deep Inelastic Scattering (DIS)}$ \\
    \end{tabular}
  \end{center}
The optimal scheme for Next to Leading Order (NLO) QCD predictions depends on 
the dominant scale. 
If the virtuality, $Q^2$, and the transverse momentum of the b~quark squared, $(p_T^{b})^2$,
are of the order of the b mass squared, $m_b^2$, threshold effects due to the b mass 
need to be considered and the so called massive scheme (used in the Fixed 
Flavour Number Scheme - FFNS) is used. Once $Q^2$ or $(p_T^{b})^2$ are 
much larger 
than $m_b^2$, massless schemes (used in the Zero-Mass Variable 
Flavour Scheme - ZM-VFNS) can be used. Here the b quark is treated 
as a massless parton in the hard scatter. In the Variable Flavour Number 
Scheme (VFNS) the massive and massless approaches are combined.
Differential cross 
sections are compared to massive NLO predictions by FMNR
for $\gamma$p and 
HVQDIS
for DIS. For references to all schemes, NLO predictions and MC generators
please refer to ref.~\cite{Aktas:2004az,Aktas:2005zc,Chekanov:2003si,Chekanov:2004tk}.
\section{Measurements in $\mu$+jet(s)\label{sec: pphp}}
In events with a jet and a muon, two properties of the b hadrons were exploited
to tag beauty events:
the large b mass and the long lifetime. The b mass leads to a relatively large
transverse momentum of the muon relative to the closest jet, 
$p_T^{\mathrm{rel}}$. 
The long lifetime of the b was used by measuring the distance of closest 
approach in the transverse plane, $\delta$, of the muon track to the beam 
spot or main vertex. 
A sign was added to $\delta$, which was set positive if the muon track crosses 
the jet axis downstream of the beam spot,
else it was set negative. 
\subsection{Beauty in $\gamma$p in $\mu$+jets}
\begin{figure}[thb]
  \begin{center}
    \begin{picture}(120,57)
      \put(0,-3){\epsfig{file=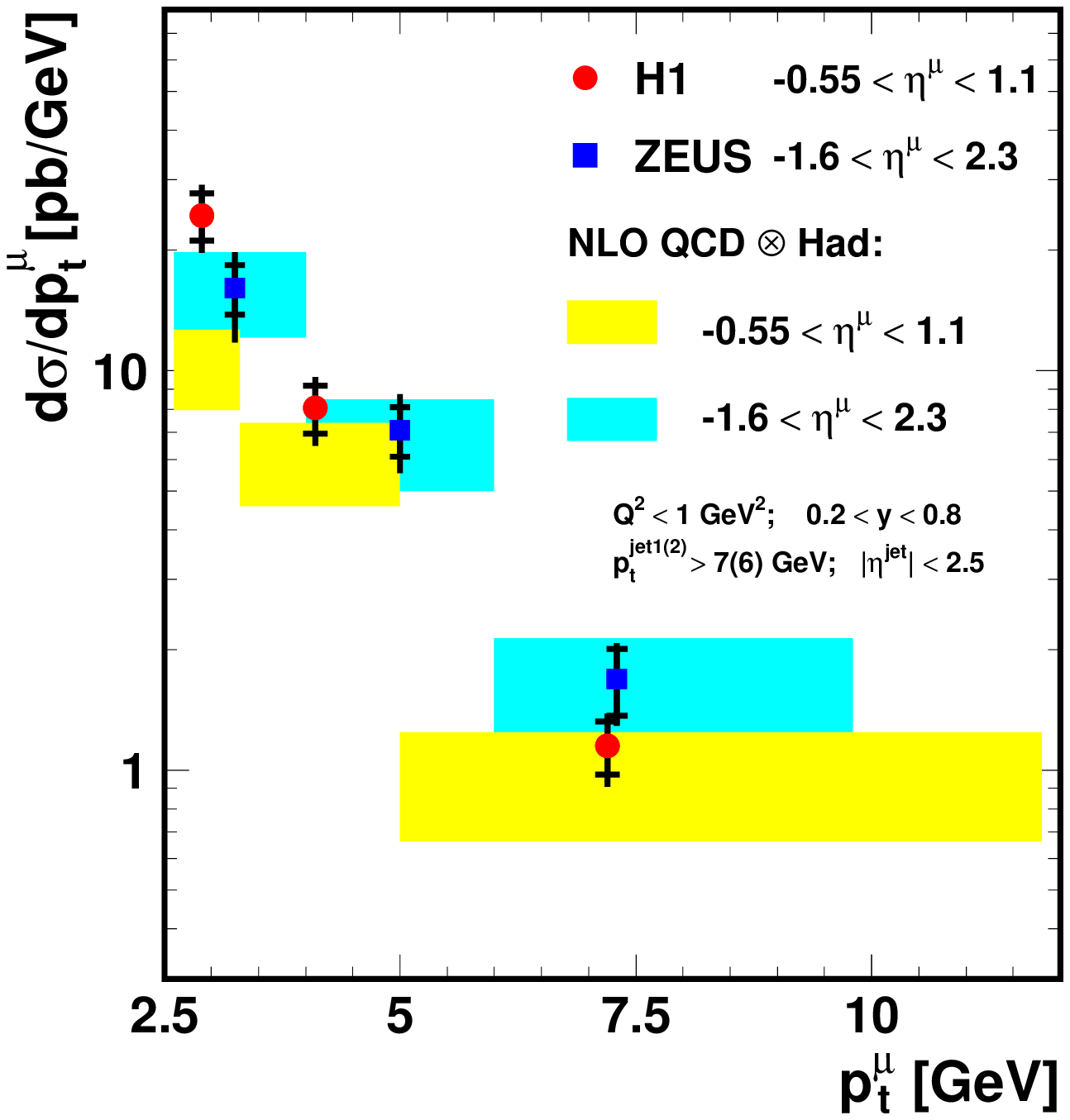, width=6cm, clip}}
      \put(50,7){(a)} 
      \put(60,-3){\epsfig{file=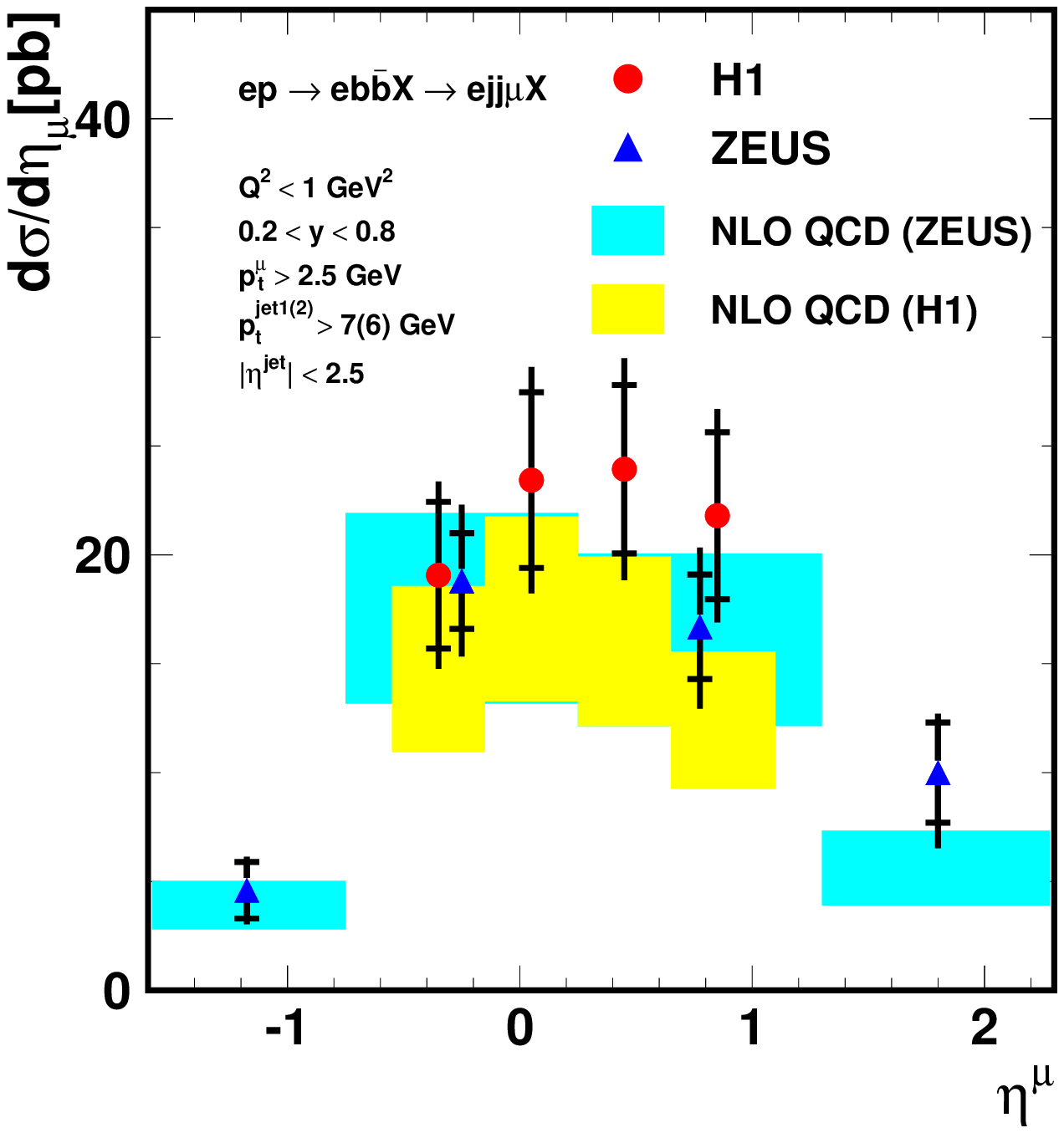, width=6cm, clip}}
      \put(110,7){(b)} 
    \end{picture}
  \end{center}
  \caption{Differential cross sections in $p_T^{\mu}$
    and $\eta^{\mu}$ of muons from beauty events in $\gamma$p with NLO 
    predictions as measured by H1
 and ZEUS.
  }
  \label{fig: bPHP}
\end{figure}

H1 has measured differential cross sections  in events with a muon and 2 
jets~\cite{Aktas:2005zc} in the photoproduction regime for $Q^2<1$~GeV$^2$ 
using an integrated luminosity of 50~pb$^{-1}$. To tag beauty events a 
simultaneous fit to both the $p_T^{\mathrm{rel}}$ and the signed $\delta$ 
distribution was performed. 
The 
cross sections 
obtained are given in fig.~\ref{fig: bPHP}.
ZEUS performed a similar measurement~\cite{Chekanov:2003si} relying on 
$p_T^{\mathrm{rel}}$ only in a wider muon eta range of $|\eta^{\mu}| < 2.5$. \\
Differential cross sections in $p_T^{\mu}$ and $\eta^{\mu}$
are shown in fig.~\ref{fig: bPHP} together with the massive NLO QCD prediction 
(FMNR) which has been corrected, using LO+PS MC simulations, to describe the 
measured hadron level properties. 
Reasonable agreement of the NLO prediction with data is observed.
The data are a bit steeper in $p_T^{\mu}$ in the H1 measurement, see
fig.~\ref{fig: bPHP}a.
H1 and ZEUS agree within errors, as 
shown in fig.~\ref{fig: bPHP}a and b. 
\subsection{Beauty in DIS in $\mu$+jet}
\begin{figure}[htb]
  \begin{center}
    \begin{picture}(120,55)
      \put(0,-5){\epsfig{file=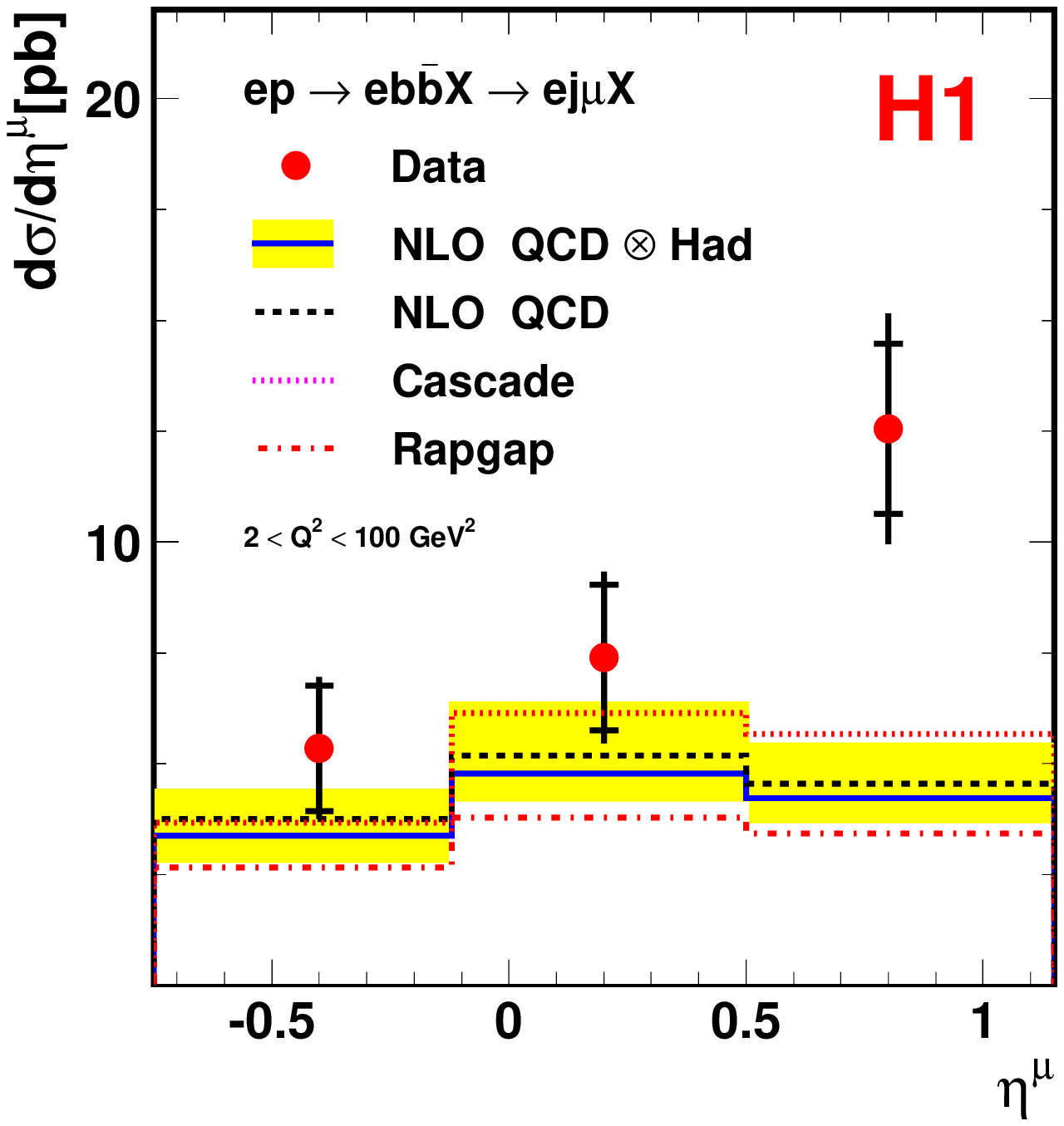, width=6cm, clip}}
      \put(50,5){(a)} 
      \put(57,-6.551){\epsfig{file=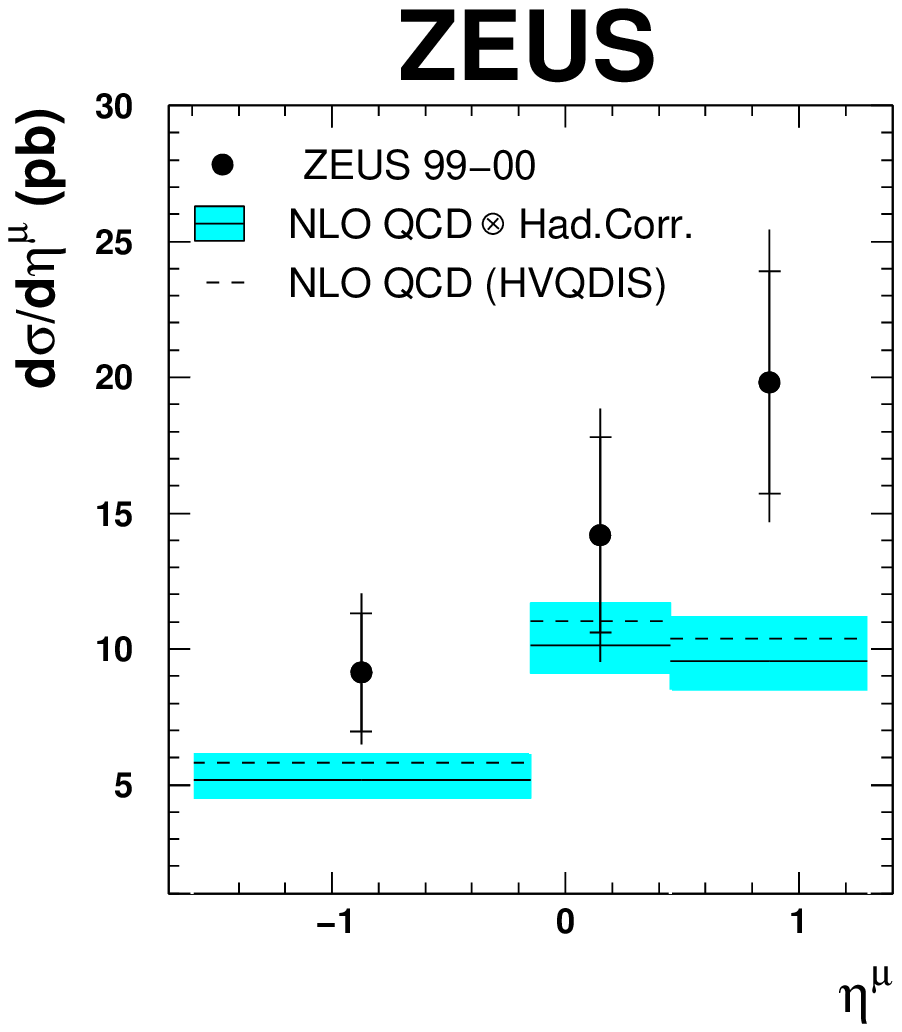, width=5.96cm, clip}}
      \put(110,5){(b)} 
    \end{picture}
  \end{center}
  \caption{The $\eta^{\mu}$ cross section of muons from beauty events in DIS
    as measured by H1 (a) and ZEUS (b). 
  }
  \label{fig: bDIS}
\end{figure}

Beauty cross sections were also measured for $Q^2>1$~GeV$^2$ using the 
same techniques as described above. H1 has 
used a $Q^2$ range of $2<Q^2<100~\mbox{GeV}^2$.
The H1 cross sections were obtained for events with
$p_T^{\mu} > 2.5$~GeV, $-0.75 < \eta^{\mu} < 1.15$, $0.1 < y < 0.7$,
$p_{T~{\mathrm{jet}}}^{\mathrm{Breit}} > 6$~GeV and 
$|\eta_{\mathrm{jet}}^{\mathrm{lab}}| < 2.5$  where 
the index 'Breit' refers to the  Breit frame. 
The $\eta^{\mu}$ cross section is shown
in fig.~\ref{fig: bDIS}a. ZEUS has performed a similar measurement in DIS
(fig.~\ref{fig: bDIS}b) using data in the $Q^2$ range of 
$1<Q^2<1000~\mbox{GeV}^2$ and defining the kinematic range as 
$p_T^{\mu} > 2.0$~GeV, $-1.6 < \eta^{\mu} < 1.3$, $0.05 < y < 0.7$, 
$E_{T~\mathrm{jet}}^{\mathrm{Breit}} > 6$~GeV and 
$-2.0 < \eta_{\mathrm{jet}}^{\mathrm{lab}} < 2.5$~\cite{Chekanov:2004tk}. 
The $p_T^{\mu}$ spectrum from both H1 and ZEUS is slightly 
steeper than the prediction (not shown). An interesting feature is seen in the
$\eta^{\mu}$ distribution in fig.~\ref{fig: bDIS}: for both H1 and ZEUS
(kinematic ranges differ)
the cross section rises towards higher $\eta^{\mu}$, 
a trend that is not well reproduced by the NLO predictions.
\section{Inclusive measurements using impact parameters}
Events with heavy flavour mesons can contain several tracks with large
impact parameters.
To inclusively determine the beauty contribution, 
events were selected in H1 by inclusive lifetime tagging, using all
$p_T > 0.5$~GeV tracks with reasonable precision vertex tracker information. 
For these tracks the impact parameter significance 
$S_{1(2)} = \delta_{1(2)} / \sigma_{\delta_{1(2)}}$ was calculated for 
the track with the largest (second largest) impact parameter $\delta$ (with 
$|\delta| < 0.1~\mbox{cm}$). The sign of the impact parameter was chosen 
relative to the direction of the closest jet if present or from the hadronic
final state, also requiring $sign(S_1) = sign(S_2)$. This method yields high
statistics and a good separation of beauty from charm and light flavour 
processes. 
\subsection{Beauty contribution to F$_2$}
\begin{figure}[thb]
  \begin{center}
    \begin{picture}(100,47)
      \put(0,-3){\epsfig{file=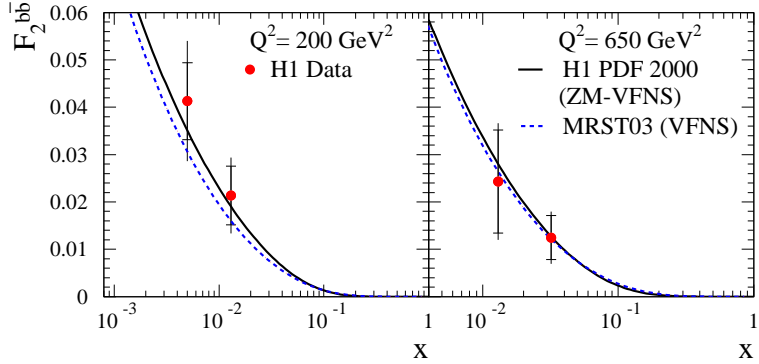, width=10cm, clip}}
    \end{picture}
  \end{center}
  \caption{The beauty contribution to the proton structure function F$_2$
  in bins of $x$ and $Q^2$. The H1 PDF 2000 fit and the MRST03 fit 
  are compared with the measurement.}
  \label{fig: f2bb}
\end{figure}

To determine the beauty contribution to the proton structure function, 
double differential beauty cross sections in bins of $x$ and $Q^2$ were 
measured using inclusive lifetime tagging. The resulting 
$\mathrm{F_2^{b\bar{b}}}$ is shown in fig.~\ref{fig: f2bb}. Good agreement 
with both the H1 PDF 2000 fit~\cite{Adloff:2003uh} and 
MRST03~\cite{Martin:2003sk} was found.
The integrated beauty cross section for $Q^2 > 150~\mbox{GeV}^2$ and $0.1 < y < 0.7$
has been determined to be $\sigma^{b\bar{b}} = 55.4 \pm 8.7 \pm 12$~pb. This is 
compatible with several predictions: 
H1~PDF~2000 (massless, {\footnotesize ZM-VFNS}) $\sigma^{b\bar{b}} = 52$~pb, 
NLO (massive$\otimes$massless, {\footnotesize VFNS}) $ \sigma^{b\bar{b}} = 47$~pb, and
NLO (massive, {\footnotesize FFNS}) $\sigma^{b\bar{b}} = 37$~pb.
\section{Conclusion and Outlook}
\begin{figure}[thb]
  \begin{center}
    \begin{picture}(140,43)
      \put(0,-3){\epsfig{file=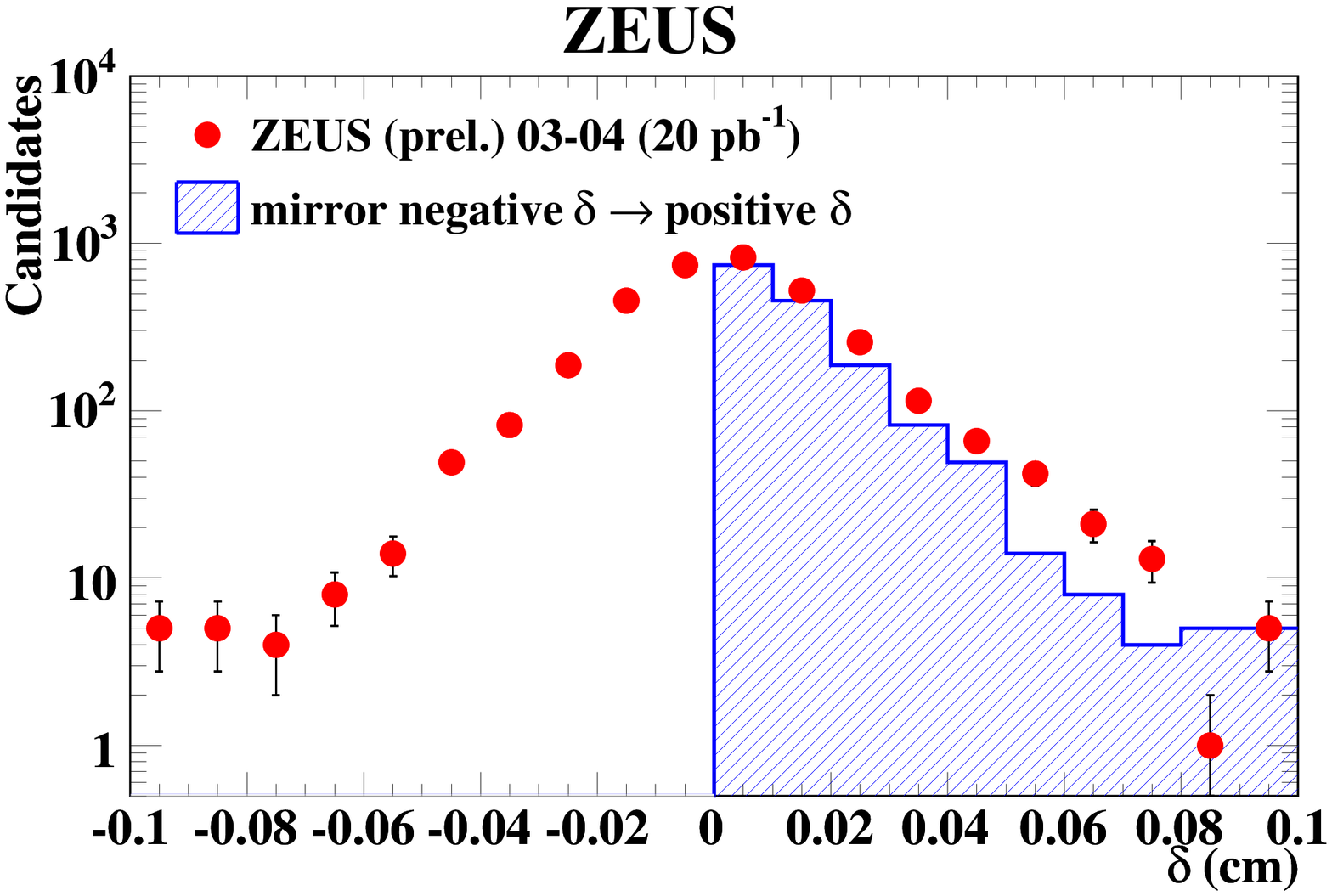, width=7cm, clip}}
      \put(60,22){(a)} 
      \put(70,-3){\epsfig{file=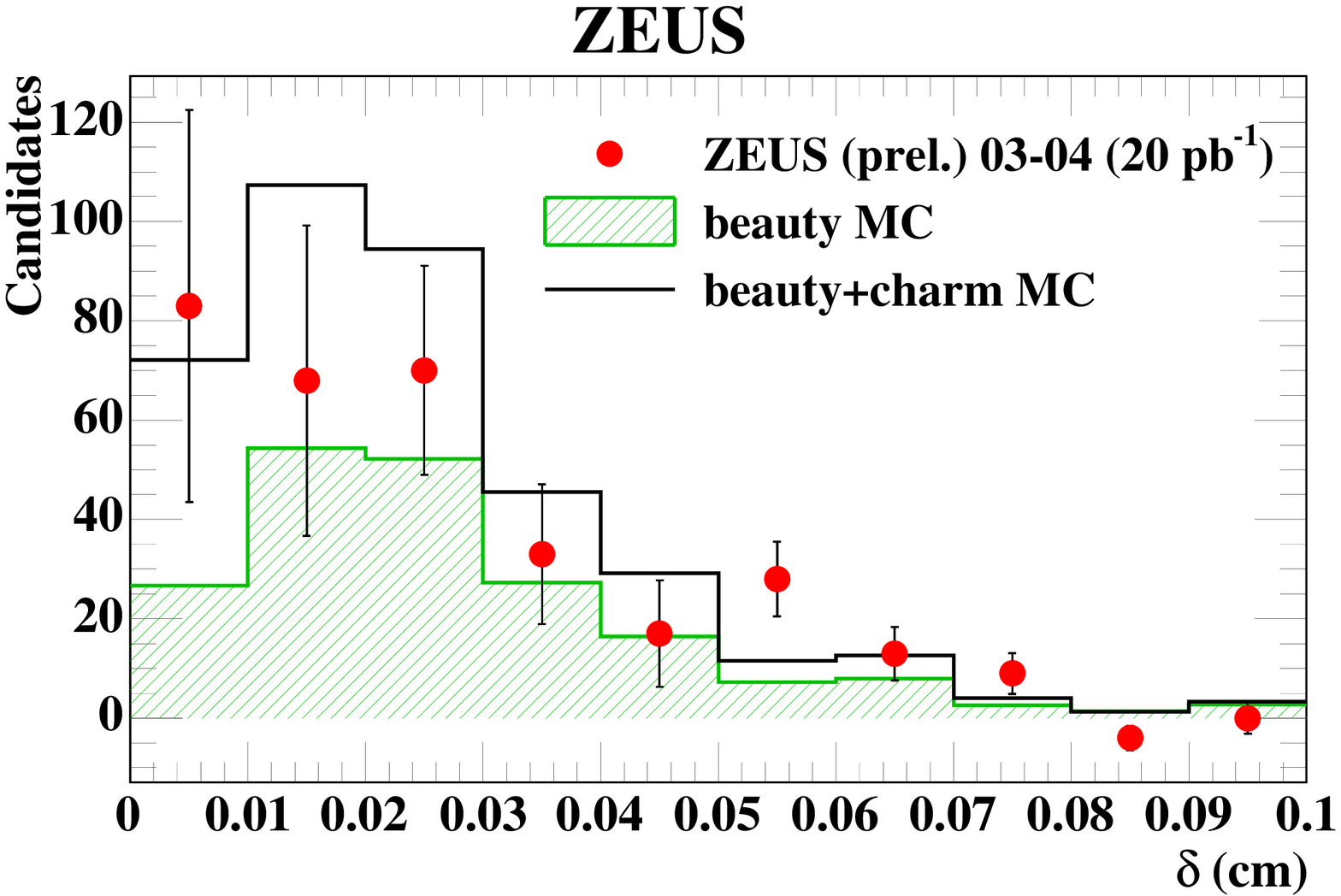, width=7cm, clip}}
      \put(130,22){(b)} 
    \end{picture}
  \end{center}
  \caption{Study of beauty in $\gamma$p using the HERA II data and impact 
    parameters measured with the new ZEUS vertex detector. (a) distribution of 
    signed impact parameters, with the mirrored negative 
    part shown on the right side in the hatched histogram below the positive 
    entries. (b) difference of
    the positive and negative part of (a). The displayed MC prediction has been
    normalised using the $p_T^{\mathrm{rel}}$ method.
  }
  \label{fig: heraII-beauty}
\end{figure}

NLO QCD predictions generally agree with the data, though they are
somewhat below the data in some kinematic regions.
H1 and ZEUS measurements with muons + jet(s) in $\gamma$p and DIS agree. 
The measurement of the beauty contribution to F$_2$ at 
high $Q^2$ using inclusive lifetime tagging is in good agreement
with NLO.\\
The upgraded HERA II detectors are giving promising results. ZEUS presented
a study of beauty in photoproduction
based on the first 20~pb$^{-1}$ 
taken in 2003/4 using the new ZEUS vertex
detector~\cite{Chekanov:2004huch}. 
The impact parameter distribution, as shown in fig.~\ref{fig: heraII-beauty}a, 
has been validated by 
comparing the difference of the positive and negative parts of the 
impact parameter distribution with MC which in turn has 
been normalised using the $p_T^{\mathrm{rel}}$ method, 
as shown in fig.~\ref{fig: heraII-beauty}b.

\end{document}